# Empirical analysis and simulation of the concave growth pattern of traffic oscillations

Junfang Tian[1], Rui Jiang[2*], Bin Jia[2], Ziyou Gao[2], Shoufeng Ma[1]

[1]Institute of Systems Engineering, College of Management and Economics, Tianjin University, Tianjin 300072, China

[2]MOE Key Laboratory for Urban Transportation Complex Systems Theory and Technology, Beijing Jiaotong University, Beijing 100044, China

This paper has investigated the growth pattern of traffic oscillations in the NGSIM vehicle trajectories data, via measuring the standard deviation of vehicle velocity involved in oscillations. We found that the standard deviation of the velocity increases in a concave way along vehicles in the oscillations. Moreover, all datasets collapse into a single concave curve, which indicates a universal evolution law of oscillations. A comparison with traffic experiment shows that the empirical and the experimental results are highly compatible and can be fitted by a single concave curve, which demonstrates that qualitatively the growth pattern of oscillations is not affected by type of bottleneck and lane changing behavior. We have shown theoretically that small disturbances increase in a convex way in the initial stage in the traditional models presuming a unique relationship between speed and density, which obviously deviates from our findings. Simulations show that stochastic models in which the traffic state dynamically spans a 2D region in the speed-spacing plane can qualitatively or even quantitatively reproduce the concave growth pattern of traffic oscillations.

**Key words:** traffic oscillation; concave growth pattern; car following behavior

## 1. Introduction

The oscillating patterns in congested traffic have been observed several decades ago (Treiterer and

* Corresponding author
Email: jftian@tju.edu.cn; jiangrui@bjtu.edu.cn.

Myers, 1974; Kühne, 1987). In order to understand the formation and evolution of traffic oscillations, various traffic flow theories and models have been proposed (see e.g., Brackstone and McDonald, 1999; Chowdhury, 2000; Helbing, 2001; Gazis, 2002; Nagel et al. 2003; Treiber and Kesting, 2013; Kerner, 2004, 2009; Saifuzzaman and Zheng, 2014; Zheng, 2014; Li et al., 2010, 2012, 2014). For instance, the formation of oscillations have been attributed to highway capacity drops (Bertini and Monica, 2005), or lane changes near the merges and diverges (Laval, 2006; Laval and Daganzo, 2006; Zheng et al. 2011). However, present interpretations about their emergence, development and propagation are still controversial (Schönhof and Helbing, 2007, 2009; Helbing et al, 2009; Treiber et al., 2010; Kerner, 2013) due to the scarcity of precise vehicle trajectory data.

Recently Jiang et al. (2014, 2015) have conducted an experimental study of car following behavior in a 25-car-platoon on an open road section. They found that the standard deviation of the vehicle velocity increases in a concave or linear way along the platoon. For the linear growth case, it can be expected that if the platoon is much longer, the variations of speed of cars in the tail of the platoon would be capped due to the physical limits of speeds. Thus the growth rate would bend downward, making the overall curve concave shaped. It has been shown that traditional car-following models based on the fundamental assumption that there is a unique relationship between traffic speed and vehicle spacing under steady state conditions contradict with this finding. They showed that by removing the fundamental assumption and allowing the traffic state to span a two-dimensional region in velocity-spacing plane, the growth pattern of disturbances has changed qualitatively and becomes qualitatively or even quantitatively consistent with that observed in the experiment.

Motivated by the experimental finding, we study the growth pattern of disturbances in NGSIM data, which were collected on real multilane highways, to check whether the concave growth pattern can be observed or not. It has been found that (1) disturbances also grow in concave way; (2) the experimental and empirical data are highly compatible, which indicates a universal evolution law of oscillations and implies that the growth pattern of oscillations is not affected by type of bottleneck and lane changing behavior.

The paper is organized as follows. Section 2 analyzes the empirical data and compares with experimental data. Section 3 analytically explains why the traditional car-following models fail to depict experimental and empirical finding. Section 4 discusses the role of stochastic factors, which are able to qualitatively change feature of the traditional models. Section 5 carries out simulations with several stochastic models that dynamically span a 2D region in the speed-spacing plane and compares with the empirical and experimental findings. Section 6 concludes the paper.

## 2. Empirical and experimental data analysis

### 2.1. Empirical data analysis

We analyzed the US-101 trajectory data in the Next Generation Simulation data sets (NGSIM, 2006), which were collected on a $640m$ segment on the south-bound direction of US-101 (Hollywood Freeway) in the vicinity of Lankershim Avenue in Los Angeles, California on June 15th, 2005. The data collected were from 07:50 a.m. to 08:35 a.m. Figure 1 provides a schematic illustration of the location for the vehicle trajectory datasets. There are five main lanes throughout the section, and an auxiliary lane is present through a portion of the corridor between the on-ramp and off-ramp. In order to minimize the impact of bottlenecks on traffic flow, only the leftmost lane is analyzed. The following criteria are used to filter suitable oscillations from the empirical data:

1. The formation process of oscillation should be comprised.
2. The speed of the leading vehicle should be as steady as possible. In other words, the standard deviation of the speed of the leading vehicle should be as small as possible.
3. Trajectories involved in jams propagating from downstream section of the investigated area were abandoned.
4. Vehicles that conducted lane changing are excluded.

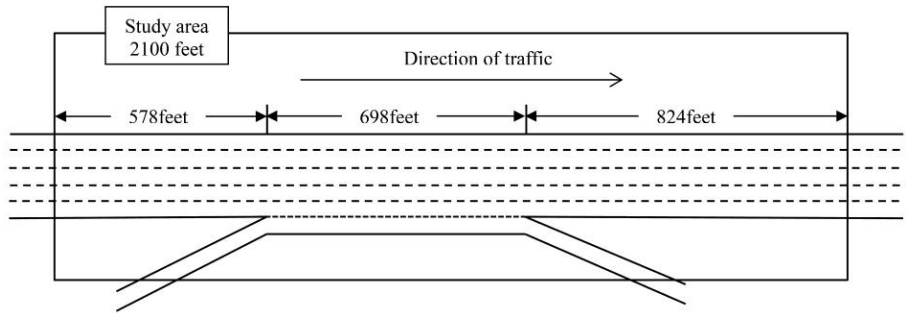

**Figure 1.** The sketch of US-101 study area.

With these criteria, 9 oscillations have been selected, see Figure 2 for two examples. Trajectories of vehicles appearing on the lane only shortly (due to lane changing) have not been considered. To understand the evolution of disturbances, standard deviations of the velocity of each vehicle have been analyzed and shown in Figure 3. For other examples, one may refer to supplemental figures in Appendix A. It can be seen that the standard deviations increase in a concave way in all examples, as observed in the car-following experiments in Jiang et al. (2014, 2015).

Figure 4(a) presents the standard deviation data in all examples in a single plot. Since the standard deviation of the speed of the leading vehicle differs from each other in different examples (see Figure 5(a)), each set of data has been shifted horizontally to make the dataset match each other (see Appendix B for an example). It can be seen that with the shifts, all datasets collapse into a single concave curve, which indicates a universal evolution law of oscillations. We note that there are deviated data marked by the ellipse. This is due to the impact of lane changing vehicles. For the vehicle indicated by red trajectory in Figure 2(a), two consecutive vehicles in its front have changed to the neighboring lane. This leads to a large acceleration of the vehicle. However, this deviation occurs only locally and does not change the global concave growth pattern.

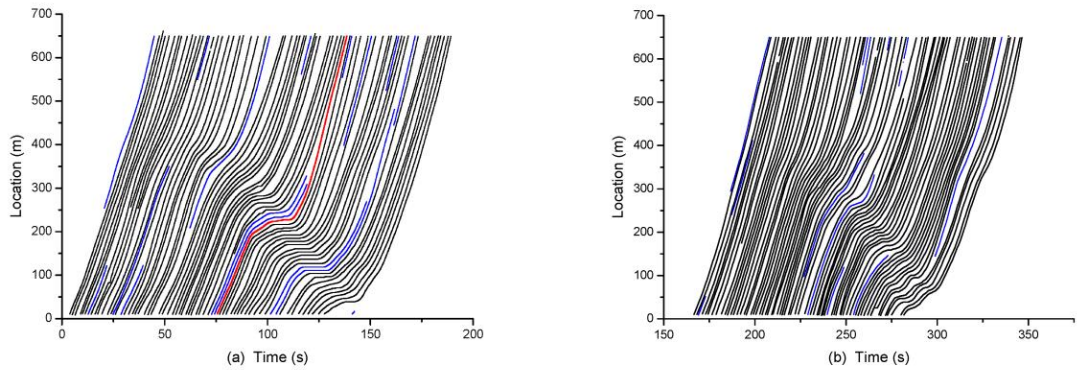

**Figure 2.** The NGSIM trajectories during the time interval 07:50 am-08:05 am. Trajectories in blue have been abandoned since vehicles only move shortly (due to lane changing) on the lane.

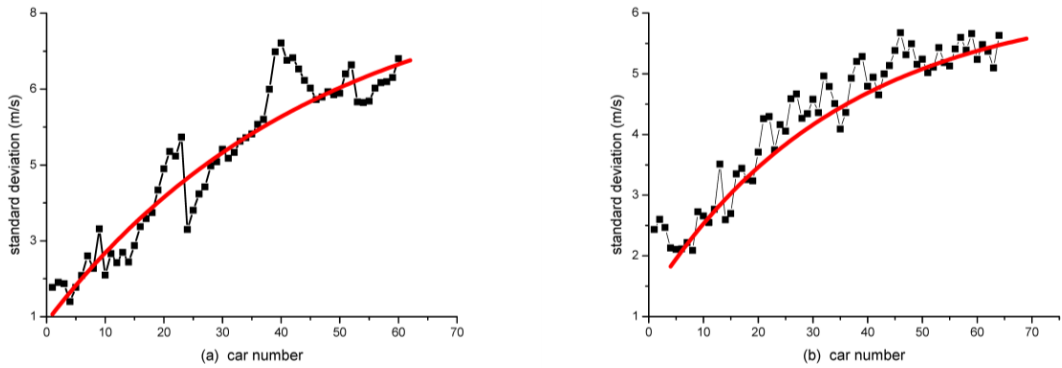

**Figure 3.** The standard deviations of the speed of each vehicle corresponding to Fig.2. The car number 1 is the leading car. The red line is a fitting line, which shows the concave growth pattern of traffic oscillations.

### 2.2. Comparison with experimental data

Now we revisit the car-following experiments in Jiang et al. (2014, 2015). In the experiments, the driver of the leading car is required to control the velocity of the car at certain pre-determined constant values $v_l$. The following 24 drivers are required to drive their cars as they normally do, but overtaking is not allowed. For each value of $v_l$, several runs of experiments have been performed.

The average speed of the leading vehicle in the 9 examples in the NGSIM data is $38 km/h$, which indicates that the oscillations can be regarded as evolving from traffic flow with average speed $38 km/h$. We present and compare the experimental results of $v_l = 40 km/h$. Note that for safety reason, the actual velocity of a car is lower than that shown by the speedometer, in particular when the velocity is high.

When the speedometer shows 40 *km/h*, the actual speed is about 38*km/h*, see Figure 5(b). For more details of traffic experiment, one may refer to Jiang et al. (2015).

Figure 6 shows the standard deviation of each car under $v_l = 40km/h$. Different from previous paper in which the average results were shown, we show the results of each run in Figure 6. One can see that for each run of the experiment, the evolution of standard deviation disperses, which is similar to the NGSIM data. However, the results under different runs can be fitted by a single concave curve.

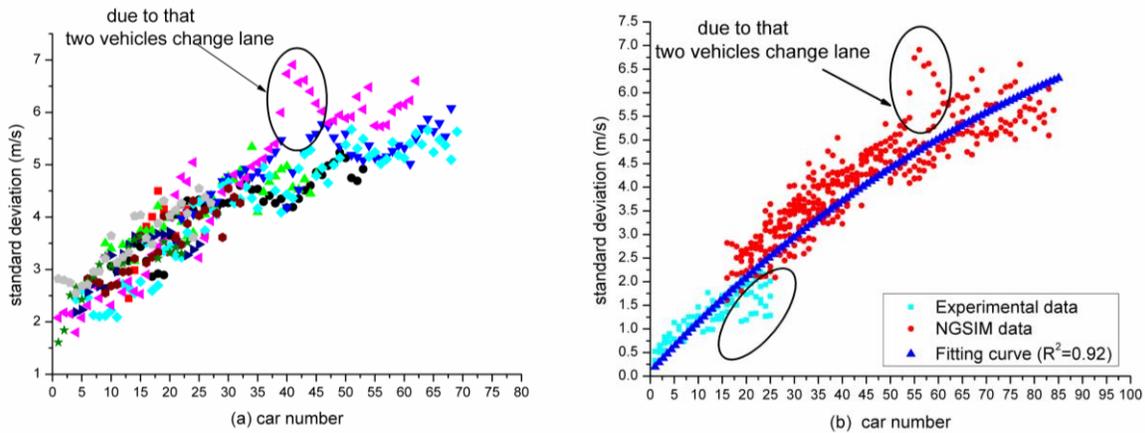

**Figure 4.** (a) The standard deviation of the speed of each vehicle in all empirical examples. (b) Merging the empirical data and the experimental data. The fitting curve is given by $y = a \exp(-x/x_0) + y_0$, where $a = -10.4$, $x_0 = 94.29$, $y_0 = 10.56$.

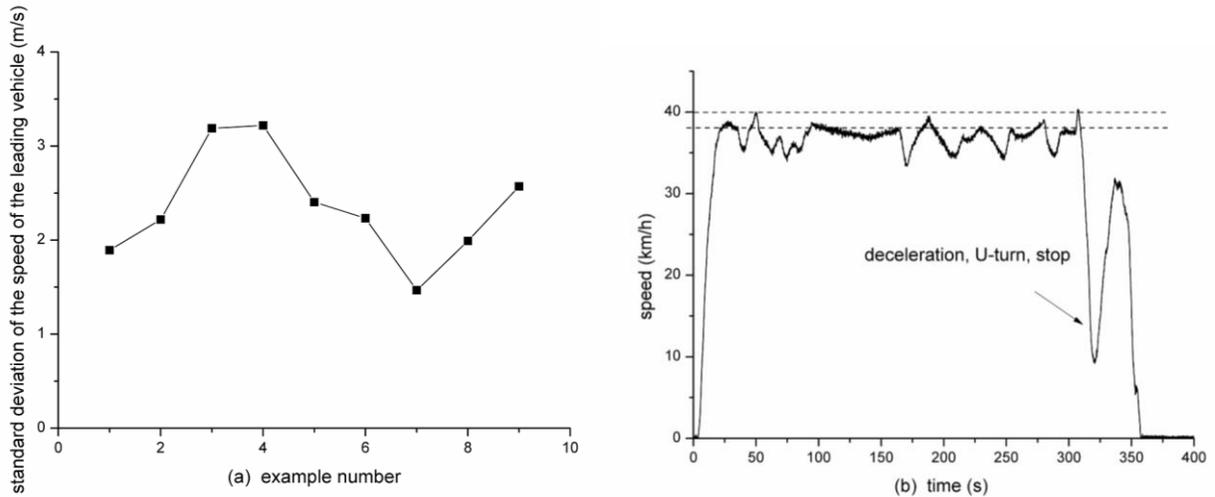

**Figure 5.** (a) The standard deviation of the leading vehicle in the 9 empirical examples. (b) The speed of leading vehicle in one run of the traffic experiment. The driver was asked to drive with 40*km/h*, but the actual speed is smaller than 40*km/h*. The two dashed lines show 40*km/h* and 38*km/h*.

Next we merge the empirical data and the experimental data in a single plot. Since the standard deviation of the speed of the leading vehicle is close to zero in the traffic experiment, we shift horizontally the empirical dataset to make the two datasets match each other, see Figure 4(b). It can be seen that the two datasets are highly compatible and can be fitted by a single concave curve. Note that the cars in the experiment were led by a moving bottleneck, and there is no lane changing. In contrast, in empirical data, traffic oscillations were induced by fixed bottleneck and vehicles can change lane. Thus, compatibility of the two datasets demonstrates that the growth pattern of oscillations is not affected by type of bottleneck and lane changing behavior.

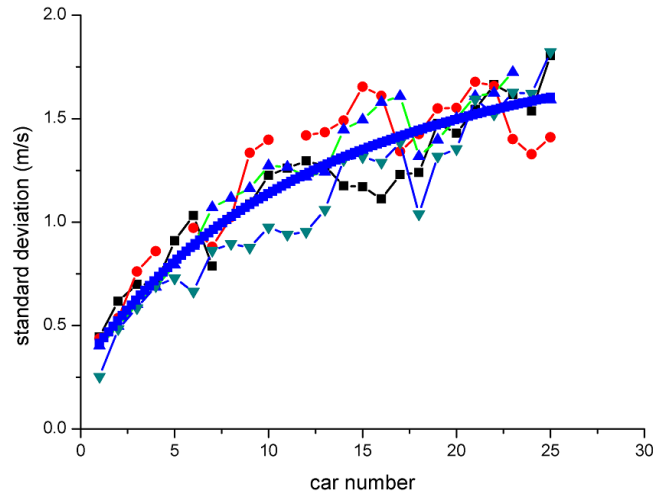

**Figure 6.** The standard deviation of the speed of each vehicle in the car following experiments. The car number 1 is the leading car. The required velocity of the leading car $v_l$ is 40*km/h*. The actual velocity is about 38*km/h*. Each symbol line represents experimental result of one run and the heavy line is the fitting curve. The discontinuity of some symbol lines is due to absence of experimental data, since the data were collected by GPS and the signal has been lost temporarily.

We notice that the experimental data marked by the ellipse slightly deviate from the fitting curve in Figure 4(b). This might be because there is no cut-in in the experiment. Therefore, some drivers leave an averagely larger spacing in front than driving in multilane road, which lowers down the growth rate of the standard deviation.

## `3. Initial convex growth pattern in traditional models

In a recent paper, Li et al. (2014) has proved that small disturbances initially grow in a convex way in nonlinear Newell model (Newell, 1961). This section generalizes the analysis to general models, which can be written as

$$\frac{d}{dt}v_n(t) = f\left(d_n(t), v_n(t), v_{n+1}(t)\right) \tag{1}$$

$$\frac{d}{dt}d_n(t) = v_{n+1}(t) - v_n(t) \tag{2}$$

Here $v_n(t)$ is the speed of vehicle $n$ and $d_n(t)$ is the spacing between vehicle $n$ and its preceding vehicle $n+1$. Different form of the acceleration function $f(d_n(t), v_n(t), v_{n+1}(t))$ leads to different models.

Assuming there is a vehicle fleet composed by $N$ vehicles on an open road and the leading vehicle $N$ runs constantly with the speed $v_0$, then the steady and homogeneous state is as follows:

$$[v_n^e(t), d_n^e(t)] = [v_0, G(v_0)] \tag{3}$$

where $d_n^e(t) = G(v_0)$ is the solution of $f(d_n^e(t), v_0, v_0) = 0$. This state implies that all the vehicles run orderly with the velocity $v_0$ and space headway $G(v_0)$, which also indicates the unique relationship between velocity and space headway. Then the linearized system around the steady state is

$$\frac{d}{dt}\overline{v_n(t)} = f^{d_n}\overline{d_n(t)} + f^{v_n}\overline{v_n(t)} + f^{v_{n+1}}\overline{v_{n+1}(t)} \tag{4}$$

$$\frac{d}{dt}\overline{d_n(t)} = \overline{v_{n+1}(t)} - \overline{v_n(t)} \tag{5}$$

where

$$\overline{v_n(t)} := v_n(t) - v_0 \tag{6}$$

$$\overline{v_{n+1}(t)} := v_{n+1}(t) - v_0 \tag{7}$$

$$\overline{d_n(t)} := d_n(t) - G(v_0) \tag{8}$$

and $f^{d_n} = \frac{df}{d(d_n)}\bigg|_{d_n=G(v_0), v_n=v_0, v_{n+1}=v_0}, f^{v_n} = \frac{df}{d(v_n)}\bigg|_{d_n=G(v_0), v_n=v_0, v_{n+1}=v_0}, f^{v_{n+1}} = \frac{df}{d(v_{n+1})}\bigg|_{d_n=G(v_0), v_n=v_0, v_{n+1}=v_0}$.

Applying the Laplace transformation to both the left-hand-side and the right-hand-side of Equation (4)

and (5), the relation between velocity disturbance of vehicle $n$ and $n+1$ can be derived

$$L(\overline{v_n(t)}) = H(s)L(\overline{v_{n+1}(t)}) \tag{9}$$

$$H(s) = \frac{f^{d_n} + sf^{v_{n+1}}}{s^2 - sf^{v_n} + f^{d_n}} \tag{10}$$

where $L$ denotes the Laplace transformation. According to the control theory, the increase of the velocity disturbance from vehicle $n+1$ to $n$ is determined by the transfer function $\|H(s)\|_\infty := \sup\{|H(j\omega)|, \omega \in [0, +\infty)\}$. Thus, when $\|H(s)\|_\infty > 1$, the amplitude of disturbance propagating from the leading vehicle $N$ to the following vehicle N-$n$ increase with ratio $(\|H(s)\|_\infty)^n$; when $\|H(s)\|_\infty < 1$, the amplitude of disturbance decays with ratio $(\|H(s)\|_\infty)^n$. Since $(\|H(s)\|_\infty)^n$ is a convex function of $n$ when $\|H(s)\|_\infty > 1$, the small disturbances always increase in a convex way in the initial stage.

Figure 7 shows the curves of the standard deviation of the vehicle speed in the optimal velocity model (OVM, Bando et al., 1995), the full velocity difference model (FVDM, Jiang et al., 2001) and the intelligent driver model (IDM, Treiber et al., 2000). The setup for the simulations can be found in Section 5. One can see that the simulated curves increase in a convex way in the front of the platoon, which is as predicted by the theoretical analysis and significantly deviates from the experimental and empirical results. It should be noted that the variations of speed in the tail of the platoon are capped due to the physical limits of speeds.

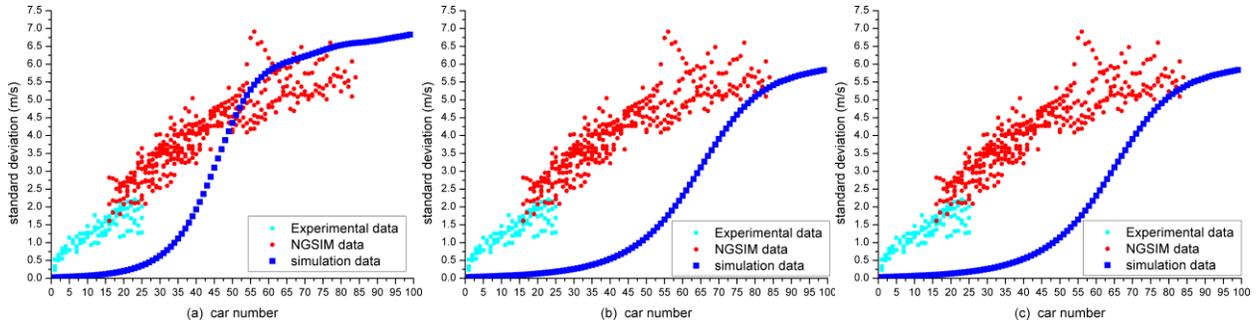

**Figure 7.** Simulation results of the standard deviation of the speed of each car. (a) the OVM, (b) the FVDM, (c) the IDM. The parameters of these three models are the same as that in Jiang et al. (2014).

## 4. Discussion

It can be seen that the traditional car following models fail to simulate the concave growth pattern of traffic oscillations. In recent years, based on the analysis of large amount of empirical data, Kerner (2004, 2009) proposed a three-phase traffic theory, which classifies congested traffic flow into synchronized flow and jam. The steady state of synchronized traffic flow is assumed to occupy a two-dimensional region in the flow-density plane. Kerner argued that in real traffic, one usually observes the phase transitions from Free flow to Synchronized flow (F→S transition) and from Synchronized flow to Wide moving jams (S→J transition). Thus, the transition from free flow to jams usually corresponds to a F→S→J process.

Later, Tian et al. (2016) have generalized Kerner's three-phase traffic theory. Since there always exist perturbations in traffic flow, the steady state cannot be realized. As a result, it has been proposed that a model is potentially able to reproduce the evolution of traffic flow, provided the traffic state can dynamically span a 2D region in the speed-spacing plane.

Taking Newell model (Newell, 2002) as an example. Figure 8(a) shows the speed of a following vehicle, in which the leading vehicle moves with constant speed. One can see that in Newell model without stochastic term, the speed and spacing of following vehicle are constant (Figure 8 (a)). Therefore, the state corresponds to a single point in the speed-spacing plane (Figure 8 (c)). On the other hand, when a stochastic term is considered (Laval et al., 2014) and is large enough, the speed and spacing of the following vehicle change significantly with time (Figure 8 (b)). The state thus dynamically spans a 2D region in the speed-spacing plane (Figure 8 (d)). In other words, the introduction of the strong stochastic term has qualitatively changed feature of the Newell model, as allowing desired time headway to change with time has qualitatively changed feature of the IDM (Jiang et al., 2014), or introducing randomization has qualitatively changed feature of the Nagel-Schreckenberg model (Nagel and Schreckenberg,1992). In the following section, it can be seen that these changes can make models reproduce the concave growth pattern of traffic oscillations.

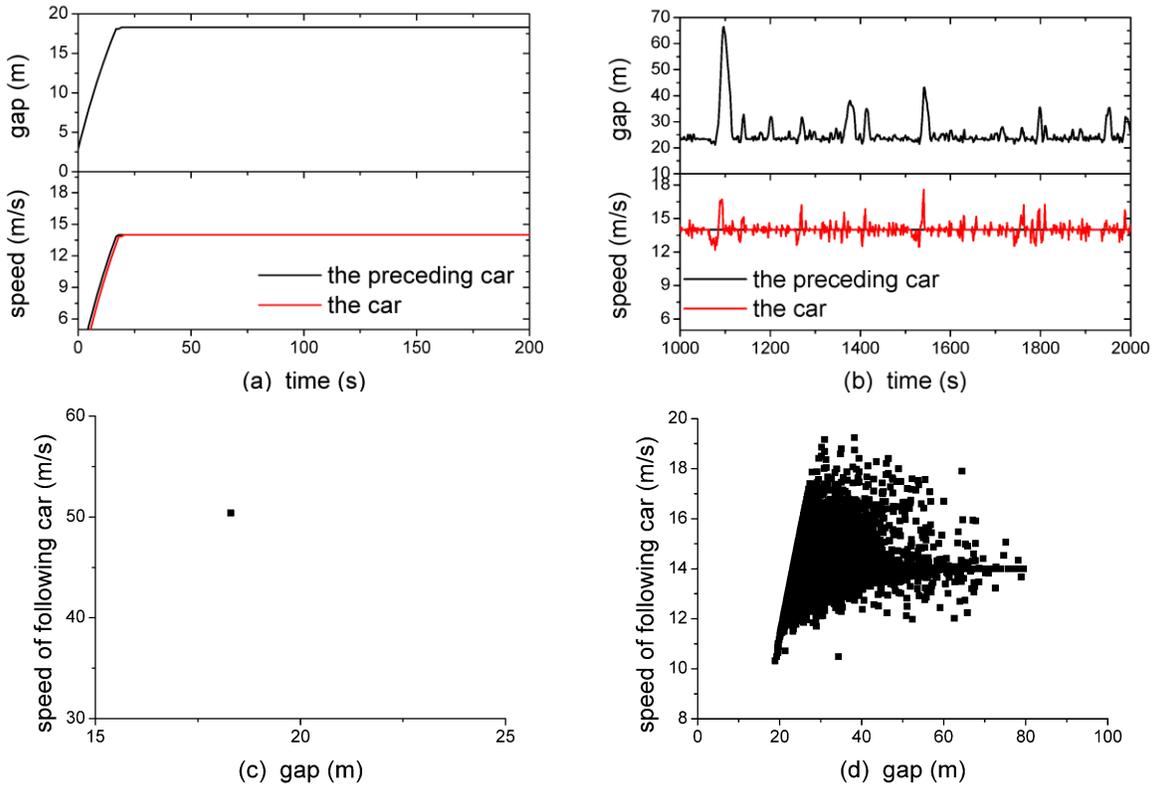

**Figure 8.** The evolution of speeds of a car and its preceding car and their gap. (a, c) Newell model (Newell, 2002) and (b, d) Newell model with stochastic term (Laval et al., 2014). The model and its parameters can be seen in Section 5.2.

## 5. Simulations

This section carries out simulations of several microscopic models, which are potentially able to reproduce the growth pattern of disturbances because traffic state in these models spans a 2D region. We compare simulation results with the empirical and experimental ones to examine these models. In the simulations, we study a platoon of 95 vehicles. All vehicles are initially stopped with their standing spacing and the leading vehicle accelerates freely to a prescribed velocity $v_l$ and then moves constantly with $v_l$. In the simulations, $v_l$ is set to $38km/h$ to match with empirical and experimental data.

### 5.1. The 2D-Intelligent Driver model

The 2D-Intelligent Driver Model (2D-IDM, Jiang et al., 2014) is given by

$$a_n(t) = a_{max}\left(1-\left(\frac{v_n(t)}{v_{max}}\right)^4-\left(\frac{v_n(t)T(t)+d_0-\frac{v_n(t)\Delta v_n(t)}{2\sqrt{a_{max}b}}}{d_n(t)}\right)^2\right) \quad (11)$$

$$T(t+\Delta t) = \begin{cases} T_1 + rT_2 & \text{with probability } p, \\ T(t) & \text{otherwise.} \end{cases} \quad (12)$$

where $v_{max}$ is the maximum velocity, $a_{max}$ is the maximum acceleration, $b$ is the comfortable deceleration, $d_0$ is the jam gap. $d_n(t)$ is the spacing between vehicle $n$ and its preceding vehicle $n+1$, $d_n(t) = x_{n+1}(t)-x_n(t)-L_{veh}$, $x_n(t)$ and $v_n(t)$ are the position and speed of vehicle $n$. $\Delta v_n(t)$ is the velocity difference between vehicle $n$ and $n+1$. $L_{veh}$ is the length of the vehicle. In contrast to the IDM where $T$ is a constant parameter, the desired time gap $T(t)$ of the 2D-IDM is a stochastic quantity that might change its value in each simulation step $\Delta t=0.1s$, where $r$ is uniformly distributed random number between 0 and 1. The parameters $T_1$ and $T_2$ indicating the range of time gap variations give rise to two-dimensional flow-density data in congested states.

In the simulation, the parameters are calibrated as: $v_{max}=120km/h$, $a_{max}=0.6m/s^2$, $b=2.0m/s^2$, $d_0=1.5m$, $T_1=0.5s$, $T_2=1.9s$, $p=0.015$, $\Delta t=0.1$s, and $L_{veh}=5m$. Figure 9(a) shows that the formation and evolution of oscillation looks very similar to that in NGSIM data. Figure 9(b) shows that the growth rate of the disturbances is also consistent with the empirical and experimental data.

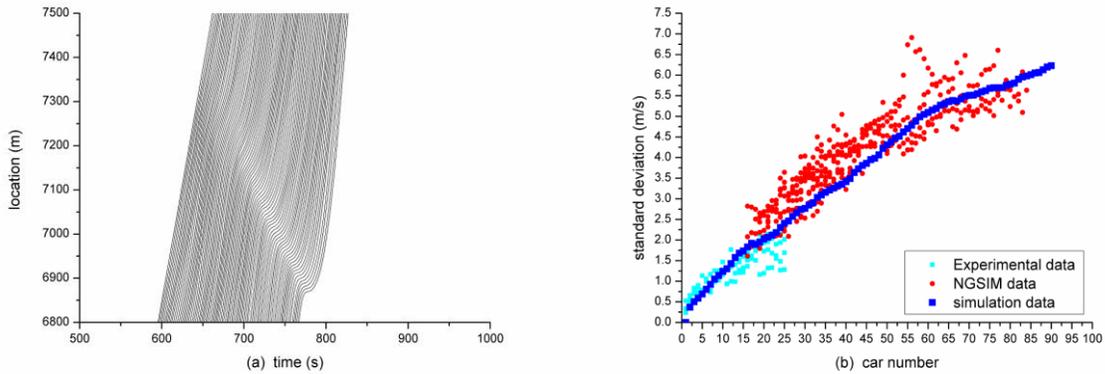

**Figure 9.** (a) The simulation results of trajectories by 2D-IDM and (b) the corresponding standard deviation of velocities of vehicles.

### 5.2. The Newell model with stochastic term

Laval et al. (2014) proposed the Stochastic Desired Acceleration Model (SDAM) in the framework of Newell's car-following model (Newell, 2002) to illustrate the formation and propagation of traffic oscillations in the absence of lane changes by adding a white noise to driver's desired acceleration. The position and velocity of vehicle $n$ is calculated by

$$x_n(t) = \min\left(x_n(t-\tau) + \max\left(\min(\xi_n(t), v_{\max}\tau), 0\right), x_{n+1}(t-\tau) - \delta\right) \tag{13}$$

$$v_n(t) = \frac{x_n(t) - x_n(t-\tau)}{\tau} \tag{14}$$

where $\tau$ is the wave trip time between two consecutive trajectories and $\delta$ is the jam spacing, and $\xi_n(t)$ is the random velocity obtained by generating Normal pseudo-random numbers with the mean and variance determined by

$$E[\xi_n(t)] = v_{\max}\tau - (1 - e^{-\beta\tau})(v_{\max} - v_n(t-\tau))/\beta \tag{15}$$

$$V[\xi_n(t)] = \frac{\sigma^2}{2\beta^3}\left(e^{-\beta\tau}(4 - e^{-\beta\tau}) + 2\beta\tau - 3\right) \tag{16}$$

where $\beta$ is the inverse relaxation time and $\xi_n(t)$ denotes the white noise process with diffusion coefficient $\sigma^2$ with the unit of $[m^2 s^{-3}]$. The random velocity of SDAM plays the same role as the random desired time gap in 2D-IDM.

In the simulation, the following parameter values are adopted via calibration: $v_{\max}=120 km/h$, $\beta=0.03 s^{-1}$, $\sigma=0.9 m^2 s^{-3}$, $\delta=1.5 m$, and $\tau=1.2 s$. The oscillations as shown in Figure 10(a) look somewhat different from the empirical data: Some unrealistically large gaps have been observed. For example, the gap shown by the red arrow has reached $200m$. Figure 10(b) shows that although the growth rate of oscillations deviates from the empirical and experimental data to some extent, the concavity of the disturbance growth has been well reproduced.

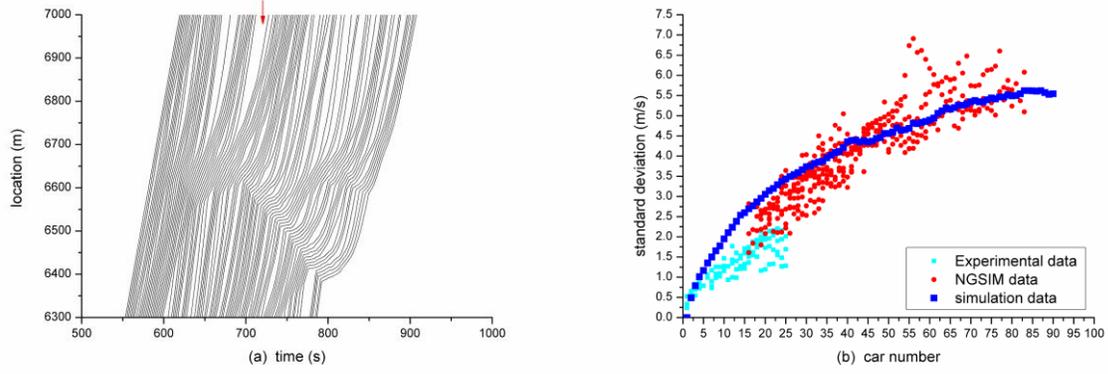

**Figure 10.** (a) The simulation results of trajectories by SDAM and (b) the corresponding standard deviation of velocities of vehicles.

### 5.3. The Kerner-Klenov-Wolf model

The Kerner-Klenov-Wolf model (KKW model, Kerner et al., 2002) is the first cellular automaton model proposed in the framework of Kerner's Three-phase theory. It includes the speed adaption effect, i.e., within the synchronized space gap, the driver will decelerate or accelerate to adapt the speed to the speed of leading vehicle. The update rules consist of the deterministic and stochastic rules.

**Step1.** Deterministic Update:

$$\tilde{v}_n(t+1) = \max(0, \min(v_{max}, v_{s,n}(t), v_{c,n}(t))).$$

where $v_{c,n}(t)$ is determined by the conditions:

$$\text{if } d_n(t) > D_n(t) - L_{veh} \text{ then } v_{c,n}(t) = v_n(t) + a,$$
$$\text{else } v_{c,n}(t) = v_n(t) + a \cdot \text{sgn}(\Delta v_n(t)).$$

where the safe speed $v_{,n}(t) = d_n(t)$.

**Step2.** Stochastic Update:

$$v_n(t+1) = \max(0, \min(\tilde{v}_n(t+1) + a\eta_n(t), \hat{v}_n(t+1))).$$

where $\hat{v}_n(t+1) = \min(v_n(t) + a, v_{max}, v_{s,n}(t))$,

$$\eta_n(t) = \begin{cases} -1 & \text{if } rand() < p_1, \\ 1 & \text{if } p_1 \leq rand() < p_1 + p_2, \\ 0 & \text{in all other cases.} \end{cases} \quad (17)$$

$$p_1 = \begin{cases} p_0 & \text{if } v_n(t) = 0, \\ p_d & \text{if } v_n(t) > 0. \end{cases} \tag{18}$$

$$p_2 = \begin{cases} p_a & \text{if } v_n(t) < v_p, \\ p_c & \text{if } v_n(t) \geq v_p. \end{cases} \tag{19}$$

**Step 3.** Car motion:

$$x_n(t+1) = x_n(t) + v_n(t+1).$$

The sign function is defined as

$$\text{sgn}(x) = \begin{cases} -1 & \text{if } x < 0, \\ 0 & \text{if } x = 0, \\ 1 & \text{if } x > 0. \end{cases} \tag{20}$$

The synchronized space gap can be defined as $D_n(t) = L_{\text{veh}} + v_n(t)h$, in which $h$ is the synchronized time gap. KKW model had incorporated the fundamental hypothesis of Kerner's Three-phase Theory into the model structure, which claims that the hypothetical steady states of the synchronized flow cover a two-dimensional region in velocity-spacing plane.

In the simulation, the following parameter values are adopted via calibration: $L_{\text{cell}}=0.5m$, $L_{\text{veh}}=15L_{\text{cell}}$, $h=2.55s$, $a=5L_{\text{cell}}/s^2$, $v_p=28L_{\text{cell}}/s$, $P_0=0.425$, $P_a=0.2$, $P_c=0.052$, $P_d=0.08$, $v_{\max}=60L_{\text{cell}}/s$. Figure 11(a) shows that the formation of oscillations can be reproduced. Figure 11(b) shows that the standard deviation increases faster in the simulation than in the empirical data in the initial stage. In particular, the standard deviation of the second vehicle has reached about $1.5m/s$, much larger than that in the experimental data. However, later the growth rate of the standard deviation in the model becomes slower than the empirical one.

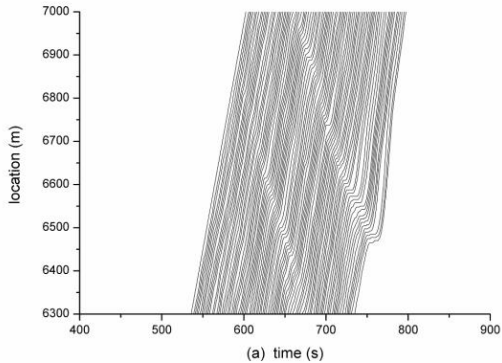
(a) time (s)

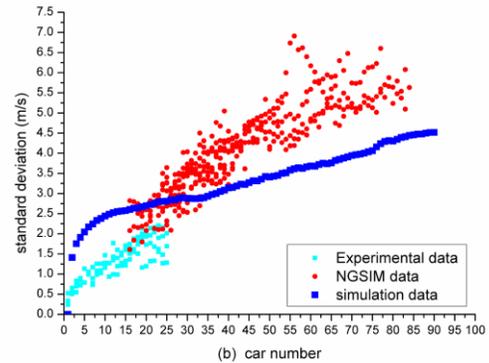
(b) car number

**Figure 11.** (a) The simulation results of trajectories by KKW model and (b) the corresponding standard deviation of velocities of vehicles.

**5.4. The Insensitive model**

In a recent paper, Jiang et al. (2015) has proposed a model based on the mechanism that in certain ranges of speed and spacing, drivers are insensitive to the changes in spacing when the velocity differences between cars are small. The model is thus named as insensitive model. It is supposed that there is a two-dimensional region R in the velocity-spacing plane. In this region, if the absolute value of the velocity difference with the preceding car $n+1$ is smaller than a threshold, i.e., $|\Delta v_n| < \Delta v_c$, the acceleration of the car changes randomly from $a_n(t)$ to

$$a_n(t+\Delta t) = \max\left(\min\left(a_n(t)+\xi, 0.1\right), -0.1\right) \tag{21}$$

in each time step $\Delta t = 0.1$. Here $\xi$ is a uniform random acceleration within the range [-0.02, 0.02]. All variables in this section are measured in SI units with time in seconds and space in meters, unless otherwise mentioned. On the other hand, if $|\Delta v_n| > \Delta v_c$, then the driver becomes sensitive to the velocity difference. In this case, the car moves as

$$\frac{dv_n}{dt} = \lambda \Delta v_n \tag{22}$$

If the state of the car is outside of region $R$ in the velocity-spacing plane, the car movement is modeled by the FVDM.

In the simulations, the parameters are calibrated as $\kappa$=0.4, $\lambda$=0.35, $v_{max}$=30, $\Delta v_c$=min(max(0.6,0.054$v_n$+0.15),1.0). The two-dimensional region $R$ is supposed to be bounded by four lines $v = 1.2(d-6)$, $v =$min(max(0.3($d$-6.8), 0.57($d$-9)), 0.35$d$+3.3), $v = v_{max}$, and $v = 0$. The OV function used is $V(d)$=max(min($v_{max}$,0.7($d$-6)),0). To avoid car collision, $v$ is set to 0 if $d \leq 6$.

Figure 12 shows that the disturbance evolution process and the trajectories of the oscillation development are generally in agreement with that of the empirical and experimental results, except that the standard deviation is a little larger than empirical results in the rear part of the platoon.

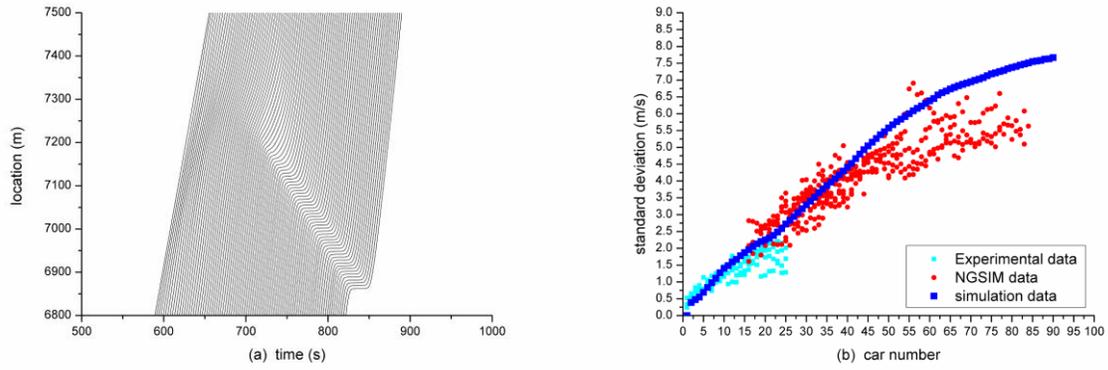

**Figure 12.** (a) The simulation results of trajectories by the Insensitive model and (b) the corresponding standard deviation of velocities of vehicles.

## 5.5 Comparison of simulation results

To further investigate which model can characterize the concave growth pattern of traffic oscillations better, we apply the error tests to quantitatively measure the goodness of fit of the standard deviations. The following two error tests are used to compare the simulation results with the field data: (1) Root mean square error (*RMSE*); (2) Root mean square percent error (*RMSPE*), which are respectively expressed as

$$RMSE = \sqrt{\frac{1}{N}\sum_{i=1}^{N}(x_i - y_i)^2} \qquad (23)$$

$$RMSPE = \sqrt{\frac{1}{N}\sum_{i=1}^{N}(\frac{x_i - y_i}{y_i})^2} \qquad (24)$$

in which *x* and *y* denote simulation results and field data, respectively. Table 1 presents the test results of vehicle standard deviation. It can be seen that 2D-IDM performs better than other models.

**Table 1.** Error tests of different models.

| MODEL | *RMSE* | *RMSPE* |
|---|---|---|
| 2D-IDM | 0.35 | 0.21 |
| SDAM | 0.47 | 0.39 |
| KKW | 1.26 | 0.51 |
| Insensitive model | 0.67 | 0.26 |

## 6. Conclusion

This paper has investigated the propagation of traffic oscillations in the NGSIM vehicle trajectories. It has been found that (1) the standard deviation of the velocity increases in a concave way along vehicle platoon in the empirical oscillations, as observed in the traffic experiments; (2) The growth pattern in different oscillations collapses into a single concave curve, which indicates a universal evolution law of oscillations; (3) the empirical data are highly compatible with the data of the car following experiment, which demonstrates that the growth pattern of oscillations is not affected by type of bottleneck and lane changing behavior.

Furthermore, we have proved theoretically that small disturbances grow in a convex way in the initial stage in the traditional car-following models presuming a unique relationship between speed and density. In contrast, simulations show that the stochastic models in which the traffic state spans a 2D region in the velocity-spacing plane can qualitatively or even quantitatively reproduce the concave growth pattern of traffic oscillations.

In our future work, we plan to conduct the following research: (1) carry out large-scale car following experiments on longer road section with larger platoon size and higher speed; (2) make efforts to collect empirical trajectory data on longer road section; (3) explore the underlying reason why models allowing 2D region can reproduce concave growth pattern; (4) study the role of traffic flow stochastic factors on the concavity of the oscillation propagation; (5) develop traffic flow models that can better reproduce the car-following features.

**Acknowledgements**

The authors wish to thank NGSIM for supplying the empirical data used in this article. JFT was supported by the National Natural Science Foundation of China (Grant No. 71401120). RJ was supported by the Natural Science Foundation of China (Grant Nos. 11422221 and 71371175). BJ was supported by the National Natural Science Foundation of China (Grant No. 71222101). SFM was supported by the National Natural Science Foundation of China (Grant No. 71431005).

## Appendix A: Supplemental Figures

This appendix shows all examples of the 9 oscillations, see Figures below.

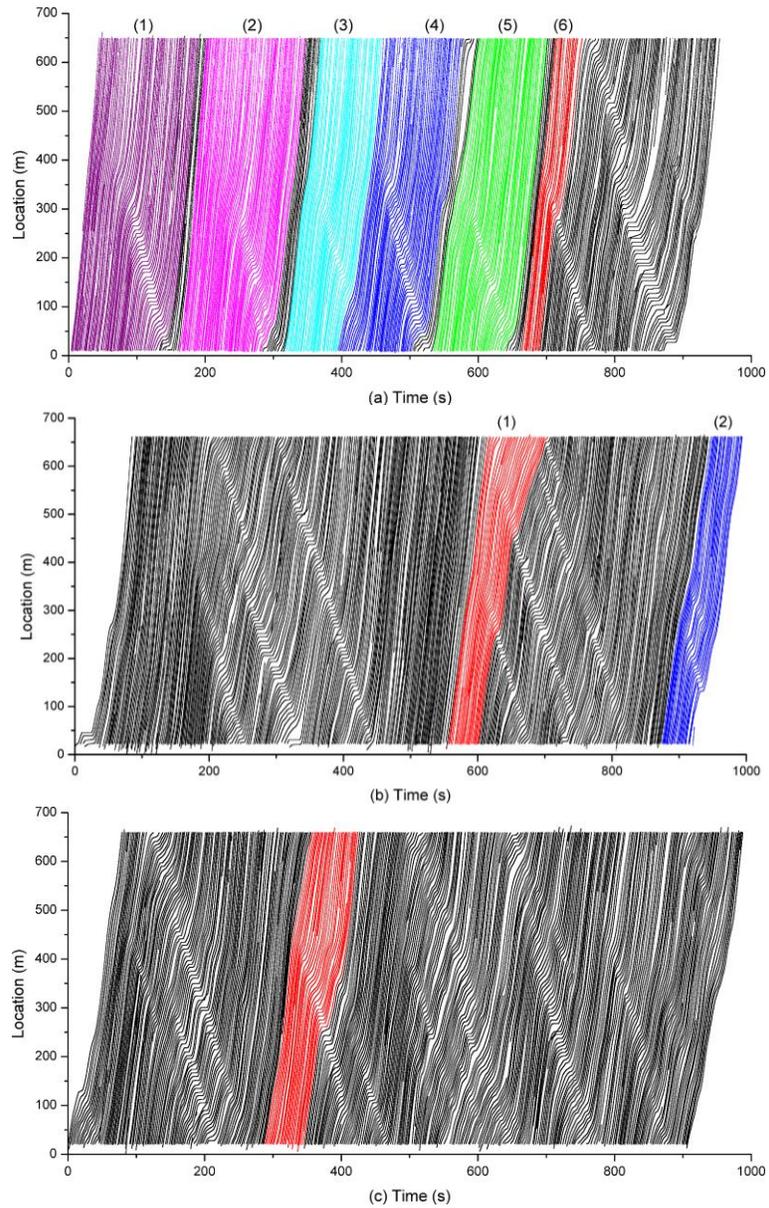

**Figure A1.** The NGSIM trajectories during the time interval (a) 07:50am-08:05am, (b) 08:05am-08:20am, (c) 08:20am-08:35am, from the leftmost lane at the location shown in Figure 1. The colored trajectories have been selected and analyzed.

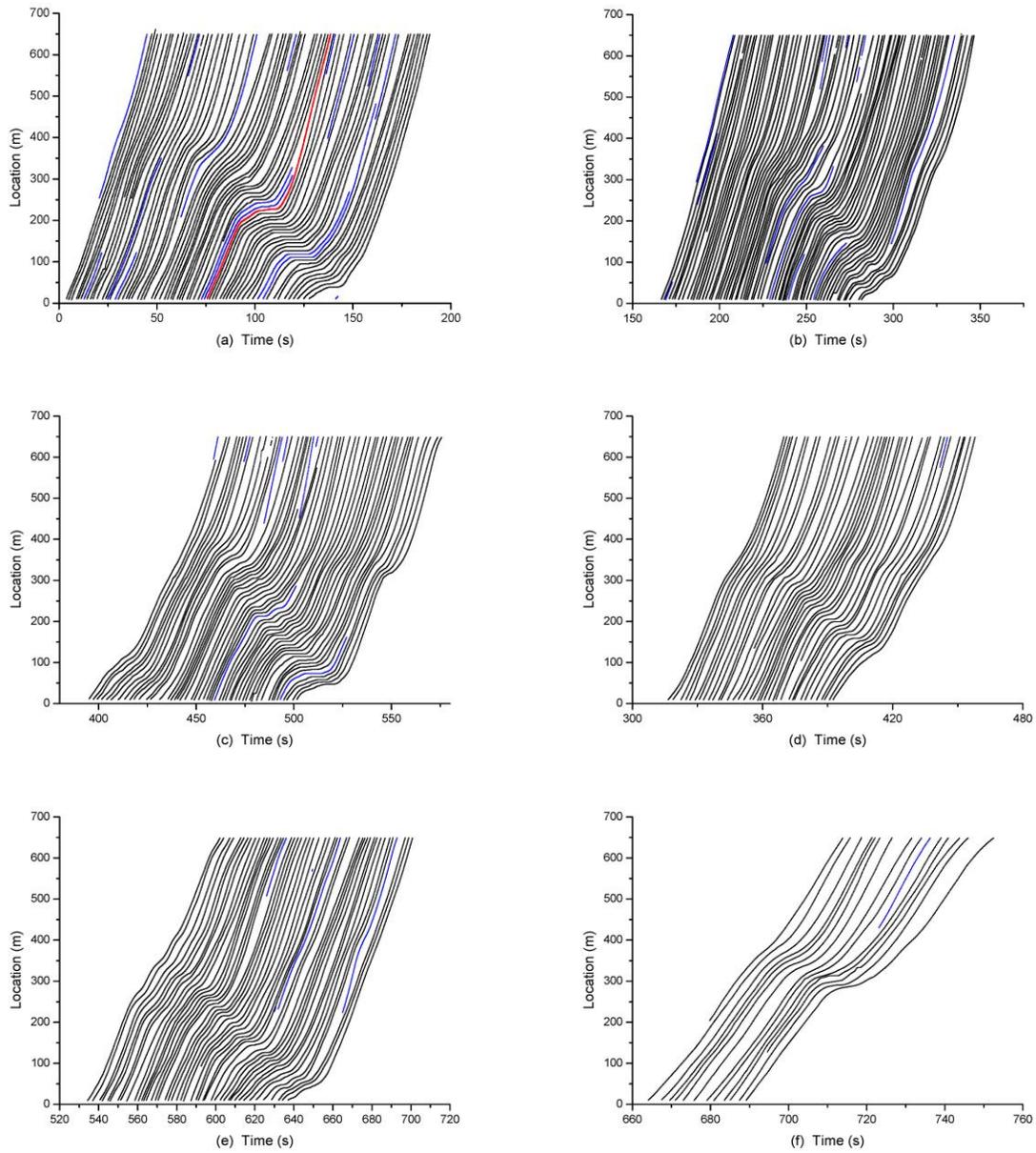

**Figure A2.** The NGSIM trajectories during the time interval 07:50 am-08:05 am. (a-f) corresponds to (1-6) in Figure A1 (a). Trajectories in black and red have been used for analysis. Trajectories in blue have been abandoned since vehicles only move shortly (due to lane changing) on the lane.

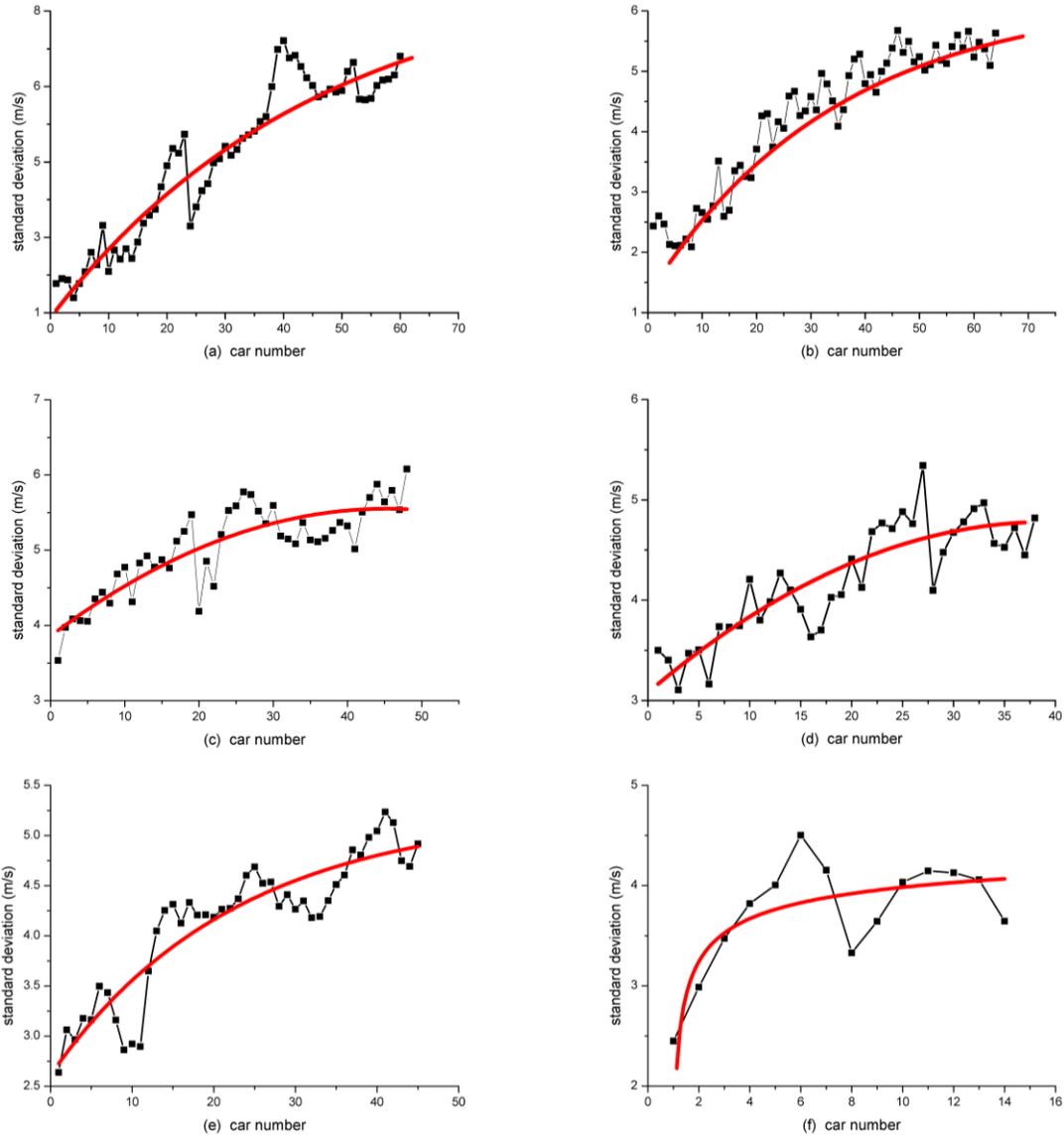

**Figure A3.** The standard deviation of the time series of the speed of each vehicle corresponding to Figure A2. The car number 1 is the leading car. The red line is a fitting line, which shows the concave growth pattern of traffic oscillations.

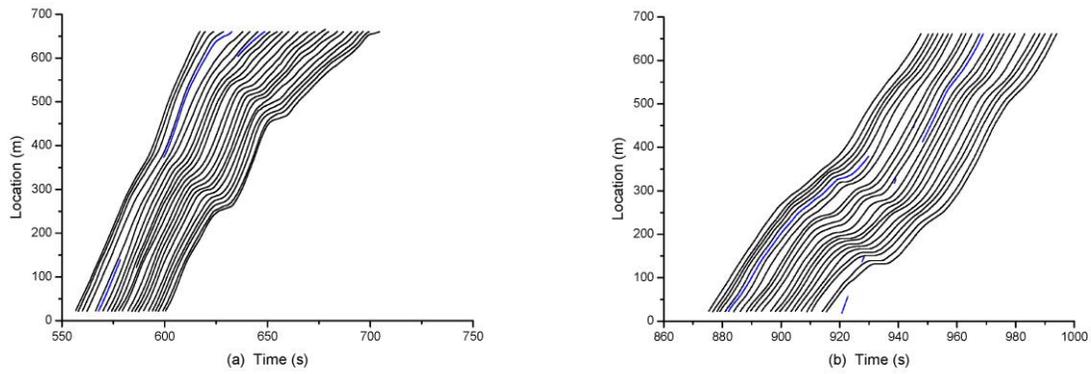

**Figure A4**. The NGSIM trajectories during the time interval 08:05 am-08:20 am. (a, b) corresponds to (1, 2) in Figure A1 (b).

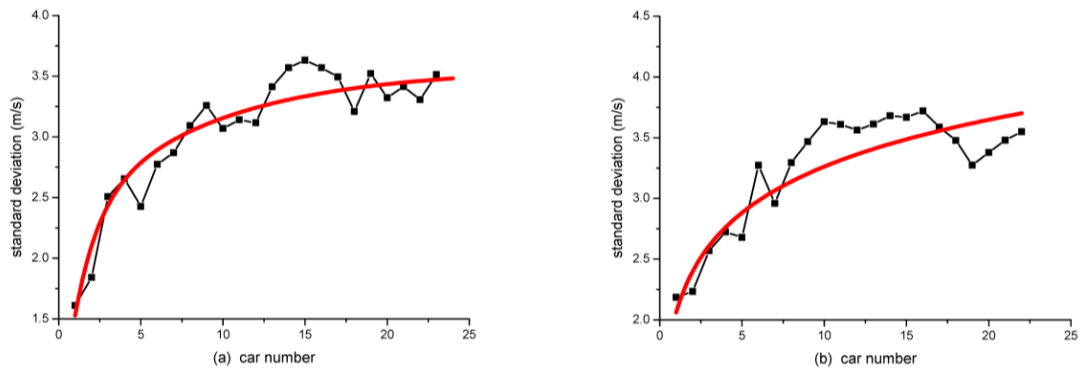

**Figure A5.** The standard deviation of the time series of the speed of each vehicle corresponding to Figure A4. The car number 1 is the leading car.

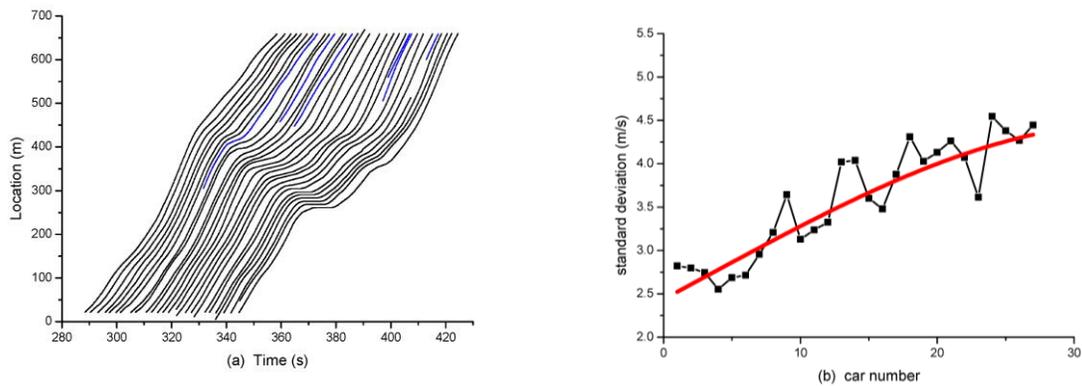

**Figure A6.** (a) The NGSIM trajectories and (b) the standard deviation of the time series of the speed of each vehicle during the time interval 08:20 am-08:35 am. (a) corresponds to the red trajectories in Figure A1(c).

**Appendix B: Example of horizontal shift of dataset**

This appendix shows an example how we horizontally shift the datasets to make them match each other, see Fig. B1. In the example, we merge two datasets: the former (black squares) corresponds to that shown in Fig.A3(b), the latter corresponds to that shown in Fig.A3(c). Since the standard deviation of the speed of the leading vehicle is much larger in the latter dataset, we have horizontally shifted the latter dataset to the right to make the two datasets match each other.

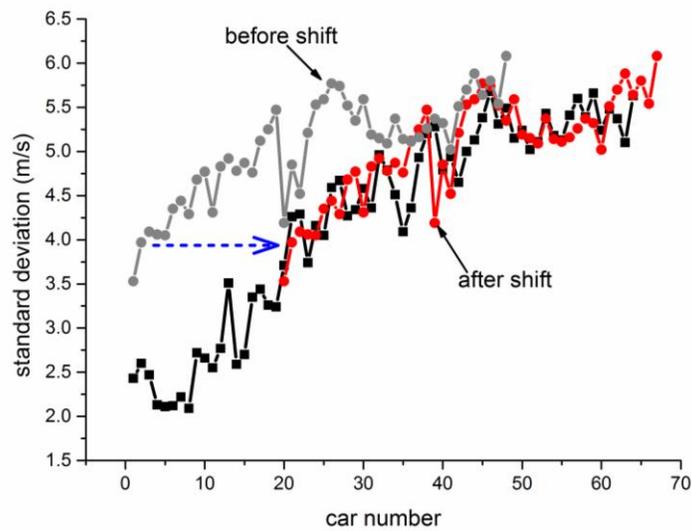

**Figure B1.** An example of horizontal shift of dataset.